\documentclass[10pt,aps,prl,reprint,nofootinbib,floats,floatfix,amsfonts,amssymb,amsmath,preprintnumbers,notitlepage,superscriptaddress]{revtex4-1}
\usepackage{aastex_macros}
\usepackage{float}
\usepackage{graphicx}
\usepackage{caption}
\usepackage{subcaption}
\usepackage[utf8]{inputenc}
\usepackage[british]{babel}
\usepackage{isomath}
\usepackage[protrusion=true,tracking=true,kerning=true,spacing=true,final,babel=true]{microtype}
\usepackage{physics}
\usepackage{amsmath}
\usepackage{amssymb}
\usepackage{xcolor}

\usepackage{url}
\usepackage[final]{hyperref}
\usepackage{tikz}

\usetikzlibrary{shapes.geometric, arrows}

\tikzstyle{startstop} = [rectangle, rounded corners, minimum width=3cm, minimum height=1cm,text centered, draw=black, fill=red!30]
\tikzstyle{io} = [trapezium, trapezium left angle=70, trapezium right angle=110, text width=3cm, minimum height=1cm, text centered, draw=black, fill=blue!30]
\tikzstyle{process} = [draw,rectangle, text width=5cm, minimum height=1cm, text centered, draw=black, fill=orange!30]
\tikzstyle{process2} = [draw,rectangle, text width=1cm, minimum height=1cm, text centered, draw=black, fill=orange!30]
\tikzstyle{decision} = [diamond, text width=2cm, text badly centered, draw=black, fill=green!30, inner sep=0pt]
\tikzstyle{arrow} = [thick,->,>=stealth]

\begin{document}

\begin{flushright}
KCL-PH-TH-2023-25
\end{flushright}

\title{The angular power spectrum of gravitational-wave transient sources as a probe of the large-scale structure}

\author{Yanyan Zheng}
\email{zytfc@umsystem.edu}
\affiliation{Institute of Multi-messenger Astrophysics and Cosmology, Missouri University of Science and Technology, Physics Building, 1315 N.\ Pine St., Rolla, MO 65409, USA}
\author{Nikolaos Kouvatsos}
\email{nikolaos.kouvatsos@kcl.ac.uk}
\affiliation{Theoretical Particle Physics and Cosmology Group, \, Physics \, Department, \\ King's College London, \, University \, of London, \, Strand, \, London \, WC2R \, 2LS, \, UK}
\author{Jacob Golomb}
\email{jgolomb@caltech.edu}
\affiliation{LIGO Laboratory, California Institute of Technology, Pasadena, CA 91125, USA}
\affiliation{Department of Physics, California Institute of Technology, Pasadena, CA 91125, USA}
\author{Marco Cavagli\`a}
\email{cavagliam@mst.edu}
\affiliation{Institute of Multi-messenger Astrophysics and Cosmology, Missouri University of Science and Technology, Physics Building, 1315 N.\ Pine St., Rolla, MO 65409, USA}
\author{Arianna I. Renzini}
\email{arenzini@caltech.edu}
\affiliation{LIGO Laboratory, California Institute of Technology, Pasadena, CA 91125, USA}
\affiliation{Department of Physics, California Institute of Technology, Pasadena, CA 91125, USA}
\author{Mairi Sakellariadou}
\email{mairi.sakellariadou@kcl.ac.uk}
\affiliation{Theoretical Particle Physics and Cosmology Group, \, Physics \, Department, \\ King's College London, \, University \, of London, \, Strand, \, London \, WC2R \, 2LS, \, UK}
\date{\today}

\begin{abstract}
We present a new, simulation-based
inference method to compute the angular power spectrum of the distribution of foreground gravitational-wave transient events. As a first application of this method, we use the binary black hole mergers observed during the LIGO, Virgo, and KAGRA third observation run to test the spatial distribution of these sources. We find no evidence for anisotropy in their angular distribution. We discuss further applications of this method to investigate other gravitational-wave source populations and their correlations to the cosmological large-scale structure.
\end{abstract}

\maketitle

\textbf{Introduction---}
Since the first detection of a gravitational wave (GW) signal from a binary black hole (BBH) coalescence in 2015 \cite{2016PhysRevLett.116.061102}, LIGO, Virgo, and KAGRA (LVK) have detected dozens more such signals during the first three observation runs \cite{GWTC3}. At the end of the next two observation runs, the number of detections is expected to reach the thousands~\cite{KAGRA:2013rdx}. 
This abundance of detected events will allow us to continuously refine our knowledge of the GW emitters.

In this context, an area of growing interest is the measurement of the spatial distribution of GW (SDGW) transient sources and its relation to the large-scale structure (LSS) of the universe~\cite{Essick:2022slj,Namikawa:2020twf,Banagiri:2020kqd,Payne:2020pmc}. The SDGW  provides a means to test the LSS that is complementary to electromagnetic measurements as well as dark siren analyses\cite{
LIGOScientific:2019zcs,LIGOScientific:2021aug}, which rely on cross-referencing GW detections with galaxy catalogs and are prone to complications such as catalog incompleteness and selection bias. Developing a scheme to accurately measure the SDGW constitutes one of the critical milestones towards precision cosmology with GWs~\cite{Baker:2019ync}.

In this paper, we present a novel, simulation-based inference method to test the SDGW that borrows from techniques used in electromagnetic precision cosmology, in particular the study of the cosmic microwave background radiation (CMB). Specifically, we show how to calculate the observed angular power spectrum of \emph{foreground} GW events and use it to probe the SDGW. This technique provides complementary information to analogous studies based on the astrophysical GW \emph{background}, where the angular power spectrum is derived from the clustering statistics of the BBH host galaxies~\cite{Jenkins:2018uac,Jenkins:2018kxc,Jenkins:2019uzp,Bertacca:2019fnt,Bellomo:2021mer}.

As a first application of our method, we test the isotropic source distribution hypothesis for the confident BBH mergers observed during the third LVK observing run (O3). However, it should be stressed that our approach is not limited to this specific instance. The technique that we present here can be easily generalized to various GW sources, future GW searches with additional detections, and different test hypotheses on the SDGW and its correlation with the LSS. 

In the next two sections, we discuss the basics of our method, the selection of GW events, the generation of synthetic signals to test the isotropic hypothesis, and the production of sky localization maps via parameter estimation. In the last two sections, we present the main results and discuss future extensions of this work.

\paragraph{\bf Methodology---}

Our method probes the spatial distribution of BBH merger events by computing their observed angular power spectrum \cite{Cavaglia:2020fnc} and comparing it to a fiducial distribution. In this work, we select the isotropic distribution, which corresponds to testing whether BBHs are isotropically distributed in the local universe. First, we compute the power spectrum of observed BBH events from the LVK GW catalogs. We choose a suitable subset of these events by imposing the selection cuts detailed in the next section. Then, we compute the power spectra of a number of mock sets obtained by injecting synthetic signals into real detector data. We sample their parameters from the latest LVK population analysis posterior distributions~\cite{LIGOScientific:2021psn} and inject the signals isotropically on the sky. We then select a subset of events by imposing the same selection cuts used for the observed BBH mergers. The synthetic power spectra are combined to produce a fiducial distribution of an isotropically distributed angular power spectrum as would be measured by the LVK detectors. Finally, we perform a statistical consistency test of the observed BBH angular distribution with the fiducial isotropic distribution; for each multipole component of the power spectrum, we compute the p-value that the observed multipole belongs to the fiducial distribution.

We consider the subset of BBH events detected during the LVK O3 observing run with a false alarm rate (FAR) smaller than 1 yr$^{-1}$ as reported in Ref.~\cite{LIGOScientific:2021psn}. We further restrict our sample to three-detector events. This is required for the generation of a consistent fiducial angular distribution, as the accuracy of sky localizations depends on the number of detectors \cite{Singer:2015ema}. These conditions restrict the sample of O3 events to 34. These events constitute our catalog of observed signals.
To generate the synthetic signals, we draw their source parameters from their inferred median population distributions \cite{LIGOScientific:2021psn}, assuming the Power Law + Peak model for the primary mass \cite{Talbot_2018} with a power law on mass ratio, the \texttt{Default} spin model \cite{Talbot2017, Wysocki2019}, and a power law model for redshift evolution \cite{Fishbach_2018}.
The phase and orientation parameters are sampled from distributions with isotropic orientations. We inject the signals into real detector data with an isotropic distribution in the sky. The times of the injections are uniformly sampled during O3. 
We then downselect these times to periods that do not overlap with known non-astrophysical transient noise \citep{O3detchar} and GWTC-3 confident detections \citep{GWTC3}.
The signals are simulated with the \texttt{IMRPhenomPv2} \citep{PhenomP, PhenomPv2} waveform model.
Selecting the synthetic events based on their FAR is computationally expensive, as it requires doing PE for the full set of events. To avoid this computational cost, we substitute the FAR selection cut with a threshold on the optimal network signal-to-noise ratio (SNR) $\rho_N$. We choose $\rho_N>10$, following the approximate threshold used for the semianalytic sensitivity estimates in Ref.~\cite{LIGOScientific:2021psn}.

We generate a catalog of 3,400 synthetic events. This allows us to produce meaningful statistical results while limiting the computational cost required to perform PE and generate the sky localization maps. We use the synthetic signals to create 100 random mock sets of 34 events each. These sets provide independent realizations of what the detectors would observe under the hypothesis that the events are isotropically distributed in the sky. We use these sets to generate the fiducial distribution.
We perform PE of all observed and synthetic events with \texttt{bilby pipe} \citep{bilby}. We use the \texttt{IMRPhenomPv2} waveform for the signal model and draw the samples from the posterior distribution with the nested sampler \texttt{dynesty} \citep{dynesty}.

We adopt the standard LVK uniform priors on the mass ratio and chirp mass from Ref.~\cite{GWTC3}. We restrict the chirp mass to a $\pm 12 M_\odot$ range around the injected values of the synthetic events and the median values of the O3 observed events. Additionally, we constrain the priors on the primary and secondary masses to be within the interval [1, 120] $M_\odot$.
The prior on all other parameters is chosen according to the uninformative priors adopted in standard LVK analyses \citep{GWTC3}. We then use the posterior samples for the declination and the right ascension to produce sky maps.

\paragraph{\bf Angular power spectrum---}

\begin{figure}
\begin{center}
\includegraphics[scale=0.31]{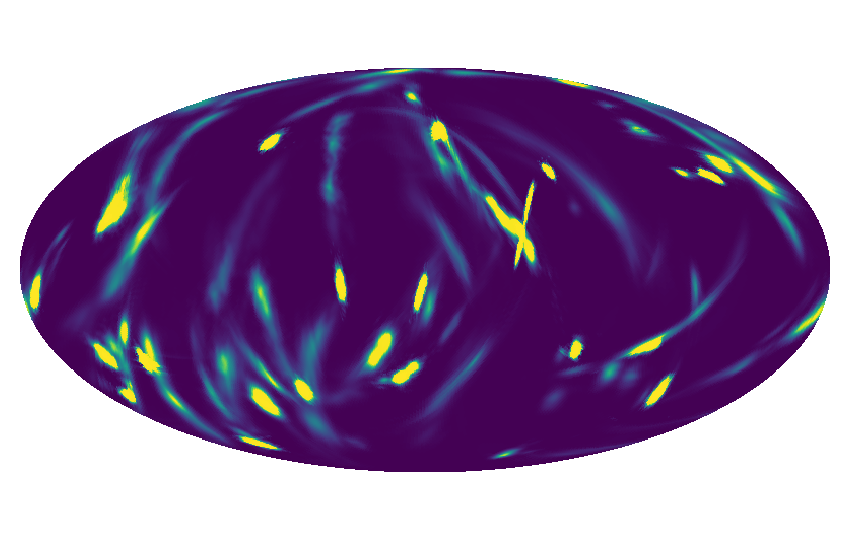}
\caption{Combined sky localization map of the O3 BBH events considered in the analysis. The sky localization of each event is generated with Bayestar \cite{Singer:2015ema} from the PE posterior samples for the declination and the right ascension. The map is created with the \texttt{Healpy} package \cite{Zonca2019,2005ApJ...622..759G}.}
\label{observed_map}
\end{center}
\end{figure}

\begin{figure}
\begin{center}
\includegraphics[scale=0.31]{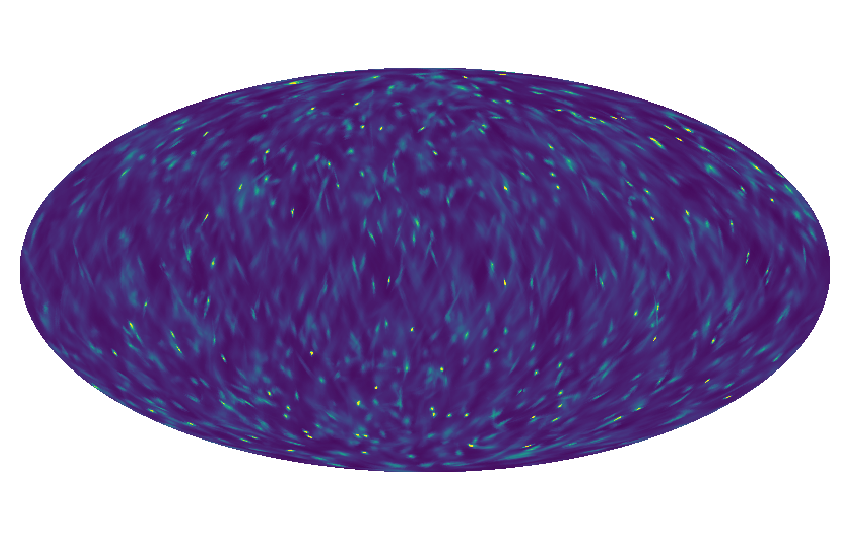}
\caption{Combined sky localization map of the synthetic BBH events that are used to build the fiducial power spectrum. Their isotropic distribution in the sky is shown by the map.}
\label{all_synthetic_maps}
\end{center}
\end{figure}

Following Ref.~\cite{Cavaglia:2020fnc}, we treat the event sky localization error regions as probability density heat maps. We generate the combined sky localization map of the observed GW events, $M(\chi,\phi)$, by stacking the sky localization density maps of all events in the observed catalog. Here, $\chi$ and $\phi$ are the polar and azimuthal angles on the celestial sphere, respectively. Figure \ref{observed_map} shows the Mollweide representation of $M(\chi,\phi)$. We repeat this procedure to obtain a cumulative sky localization map for each set of synthetic events. Figure \ref{all_synthetic_maps} shows the combined sky localization map obtained by stacking the 100 synthetic maps, each made from 34 events. The map shows that the synthetic events are isotropically distributed in the sky. It also depicts what the GW sky would look like with 3400 foreground BBH events, a not too unrealistic scenario in a few years.

We then compute the angular power spectra of the combined sky localization maps by expanding each of them into spherical harmonics:
\begin{equation}
    M(\chi,\phi)=\sum_{lm}\alpha_{lm}Y_{lm}(\chi,\phi)\,.
\end{equation}
The multipole components of the angular power spectrum are obtained by summing the absolute square of the $\alpha_{lm}$ coefficients of the expansion over $m$:
\begin{equation}
    C_l=\frac{1}{2l+1}\sum_m\abs{\alpha_{lm}}^2\,.
\end{equation}
The physical information contained in the power spectrum can also be expressed in terms of the two-dimensional (angular) correlation function (CF). The CF describes the excess probability of finding two objects in the directions $\hat n_1$ and $\hat n_2$ and angular separation $\theta$ with respect to a uniform distribution. Given the cumulative sky localization map $M(\chi,\phi)$, the CF is defined as
$C(\theta) = \langle M(\hat n_1)\cdot M(\hat n_2)\rangle_{21}$, where the average is taken over the observed sky with angular separation held fixed \cite{Cavaglia:2020fnc}. The CF can be written in terms of the power spectrum as 
\begin{equation}
C(\theta) = \frac{1}{4\pi}\sum_l (1+2l)C_lP_l(\cos\,\theta)\,,
\label{eq:CF}
\end{equation}
where $P_l(\cos\theta)$ denotes the Legendre polynomial of order $l$ and argument $\cos\theta$. Typically, the finite beam resolution of the detectors leads to a high-$l$ cutoff $l_{\rm max}$ in Eq.(\ref{eq:CF}). This effect can be modeled by introducing a window function $W_l\propto\exp[-l(l+1)\sigma_{\rm res}^2]$, where $\sigma_{\rm res}$ is the detector resolution \cite{White1994AnisotropiesIT}. 

The diffraction-limited angular resolution of the LIGO-Virgo network determines the high-$l$ cutoff as $l_{\rm max}\sim \pi/\theta_{\rm res}$, where $\theta_{\rm res}$ is the angular resolution. We estimate $l_{\rm max}$ directly from the distributions of the skymaps. We fit the distribution of the observed skymap 90\% contour regions as a proxy for the square angular resolution ${\rm\Delta}\Omega_{\rm res}=2\pi[1-\cos(\theta_{\rm res}/2)]$ with a gamma distribution. We then perform a one-tailed test and choose ${\rm\Delta}\Omega_{\rm res}$ such that 90\% of the observed events have a larger localization area than that value. This provides an estimate for the angular resolution of $\theta_{\rm res,o} \sim 6.95^\circ$, corresponding to $l_{\rm max,o}\sim 26$. We then repeat the procedure for the whole set of synthetic events. This yields $\theta_{\rm res,s} \sim 4.83^\circ$, corresponding to $l_{\rm max,s}\sim 37$. The resolution of the simulated set is better than the resolution of the observed set. We expect this is due to the larger number of events in the simulated set compared to the observations. As a consistency check, we also estimate ${\rm\Delta}\Omega_{\rm res}$ using the theoretical estimate of Ref.~\cite{PhysRevD.81.082001}. For a monochromatic GW at frequency $f$, the square angular resolution of a three-detector network is
\begin{equation}
    {\rm \Delta\Omega}_{\rm res}\approx 8 \bigg(\frac{150{\rm Hz}}{f}\frac{10}{\rho_{\rm N}}\bigg)^2\frac{10^{17}{\rm cm}^2}{A_{\rm N}}\frac{1/27}{\rho_1^2\rho_2^2\rho_3^2/\rho_{\rm N}^6}\frac{\sqrt{2}/2}{\abs{{\rm sin} {\hspace{1mm}}i_{\rm N}}}\,,
    \label{theta_res_th}
\end{equation}
where $A_N$ is the triangular area formed by the three detector sites, $i_N$ is the angle between the wave direction and the three-detector plane, $\rho_N$ is the network optimal SNR of the GW signal, and $\rho_i$ ($i=$1,2,3) are the single-detector SNRs. We consider a triangular area $A_{\rm N}=10^{17} {\rm cm}^2$ for the LIGO-Virgo network and a mean incidence angle of $45^\circ$ with the detector plane. We use the posterior sample median values to estimate the SNRs and approximate $f$ with the ISCO frequency obtained from the posterior median chirp mass and mass ratio. Using the means of the SNRs and $f$ in Eq.~(\ref{theta_res_th}), we obtain the angular resolution $\theta_{\rm res,o}\sim4.04^{\circ}$ for the observed events and $\theta_{\rm res,s}\sim4.44^{\circ}$ for the synthetic events, corresponding to $l_{\rm max,o}\sim 45$ and $l_{\rm max,s}\sim 41$, respectively. 
The theoretical estimate gives higher bounds than the data sets. This is expected, as Eq.~(\ref{theta_res_th}) is derived under optimal assumptions and a Fisher approximation. In the following, out of an excess of caution, we will use $l_{\rm max}=26$ as a conservative upper bound. 

\begin{figure}[h]
  \centering
  \includegraphics[width=0.9\columnwidth]{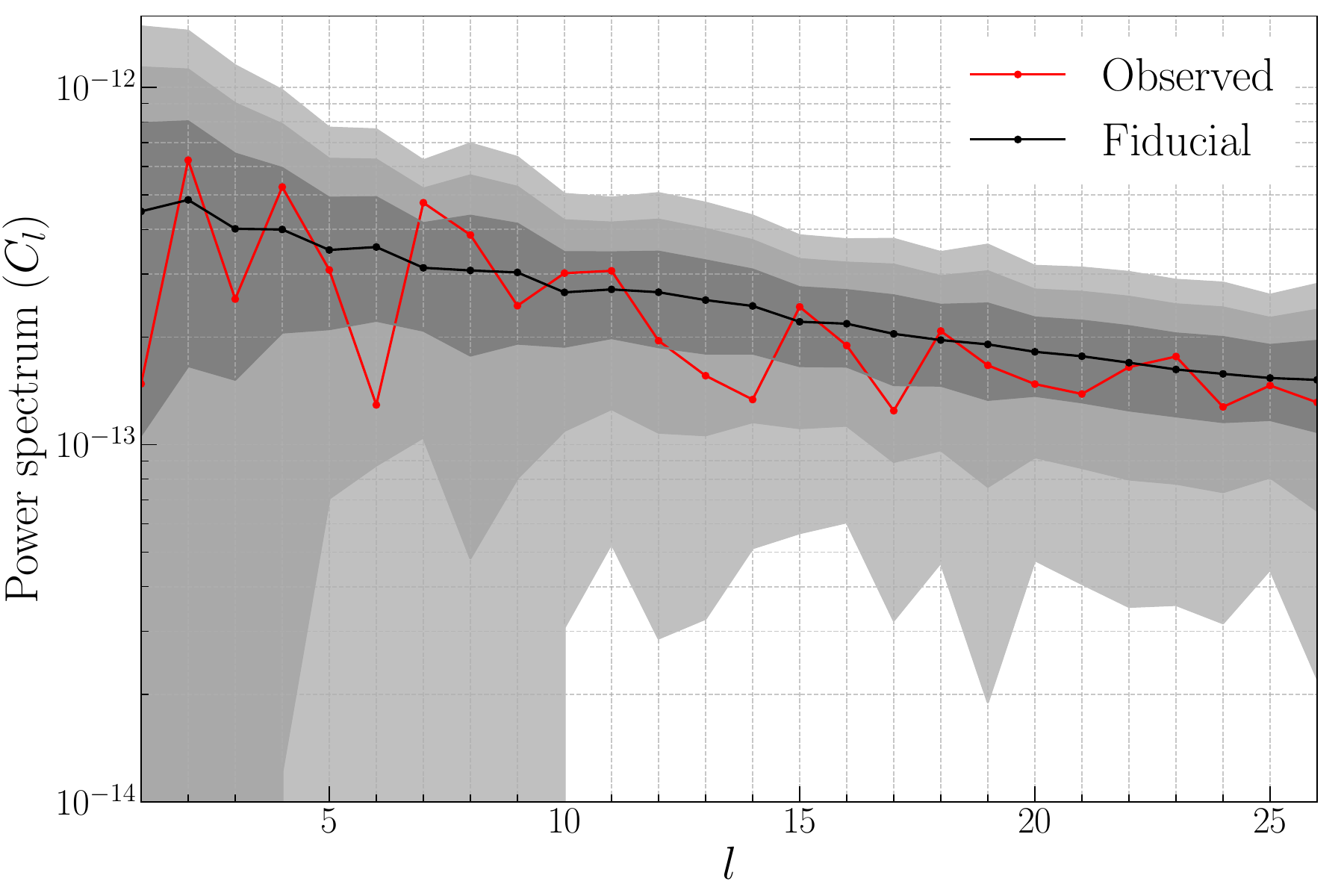}
  \captionsetup{justification=justified}
  \caption{The observed power spectrum of the O3 BBH events considered in the analysis (red curve) and the fiducial power spectrum obtained from the 100 synthetic sets under the isotropic hypothesis (black curve). The gray-filled regions denote 1 -- 3$\sigma$ deviations from the mean.}
\label{fig:cl}
\end{figure}

\paragraph{\bf Results---} Figure~\ref{fig:cl} shows the power spectrum of the observed events (red curve) and the mean spectrum of the 100 synthetic sets (black curve) up to $l_{\rm max}=26$. For each $l$, we fit the $C_l$ distribution from the synthetic sets with a gamma distribution. The three gray-filled areas in Fig.~\ref{fig:cl} (darker to lighter gray) denote the 1 -- 3$\sigma$ confidence level regions from the mean. All observed $C_l$ values lie within the 2$\sigma$ band. Therefore, we conclude that the observed angular distribution of observed BBH events shows no significant inconsistencies relative to an isotropic distribution. To quantify this statement, we performed two statistical tests. In the first test, we compute the cumulative distributions of p-values for the observed $C_l$ under the hypothesis that the BBH are distributed isotropically in the sky. 

\begin{figure}[t]
  \centering
  \includegraphics[width=0.9\columnwidth]{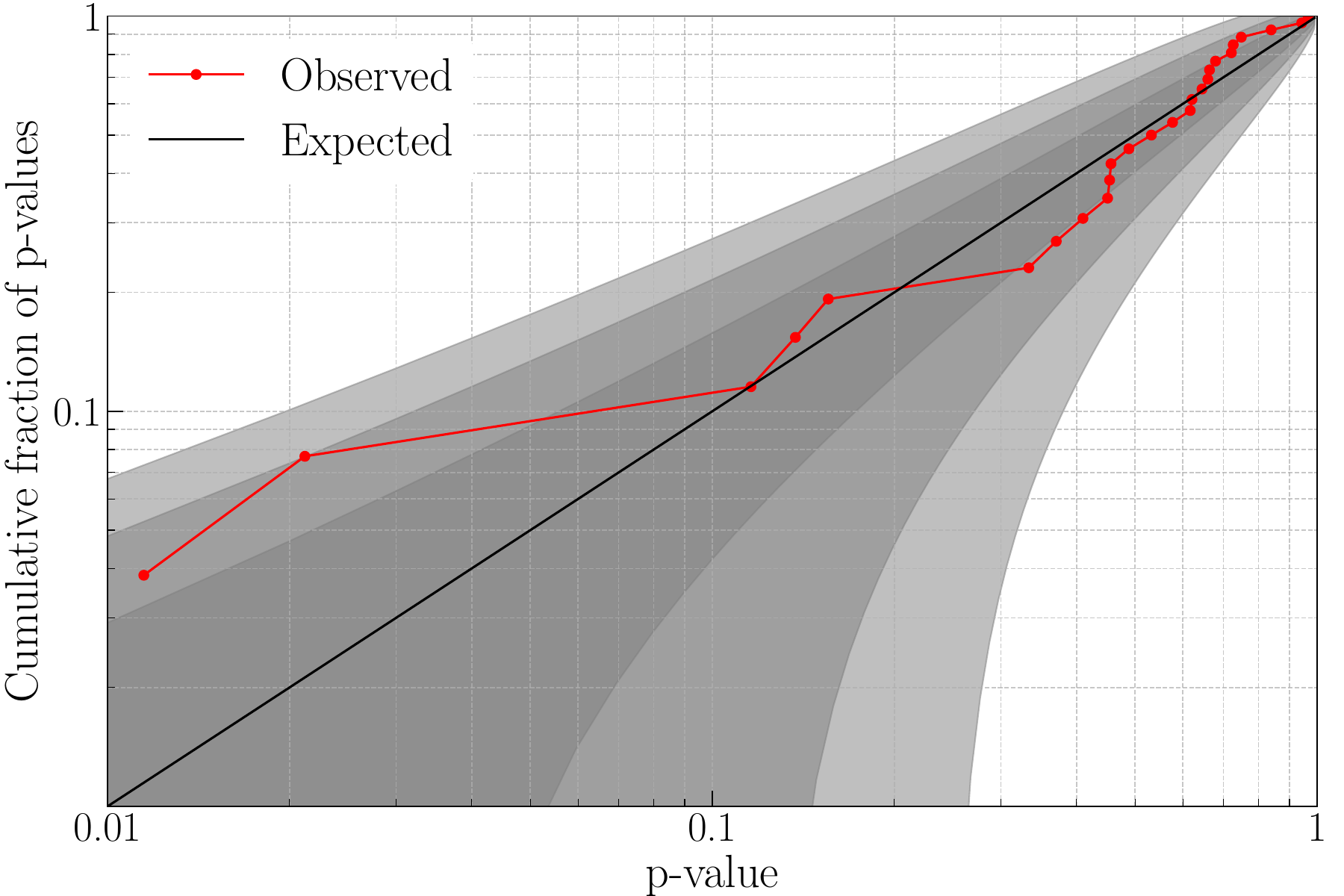}
  \caption{The cumulative distribution of observed p-values for the $C_l$. The black solid line indicates the expected distribution under the isotropic hypothesis. The gray-filled regions correspond to 1 -- 3$\sigma$ deviations from the expected distribution.}
\label{fig:pp}
\end{figure}

Figure~\ref{fig:pp} shows the cumulative distributions of p-values (red dots). The expected distribution is represented by the black dashed line, with the gray-filled regions denoting the 1 -- 3$\sigma$ confidence levels. All p-values lie within the 2$\sigma$ region, in agreement with the results of Fig.~\ref{fig:cl}. In the second test, we assess the goodness of fit of the observed power spectrum with the fiducial spectrum by performing a $\chi^2$ test, which yields a p-value of 0.82, in agreement with the null isotropic hypothesis. 

\begin{figure}[t]
  \centering
  \includegraphics[width=0.9\columnwidth]{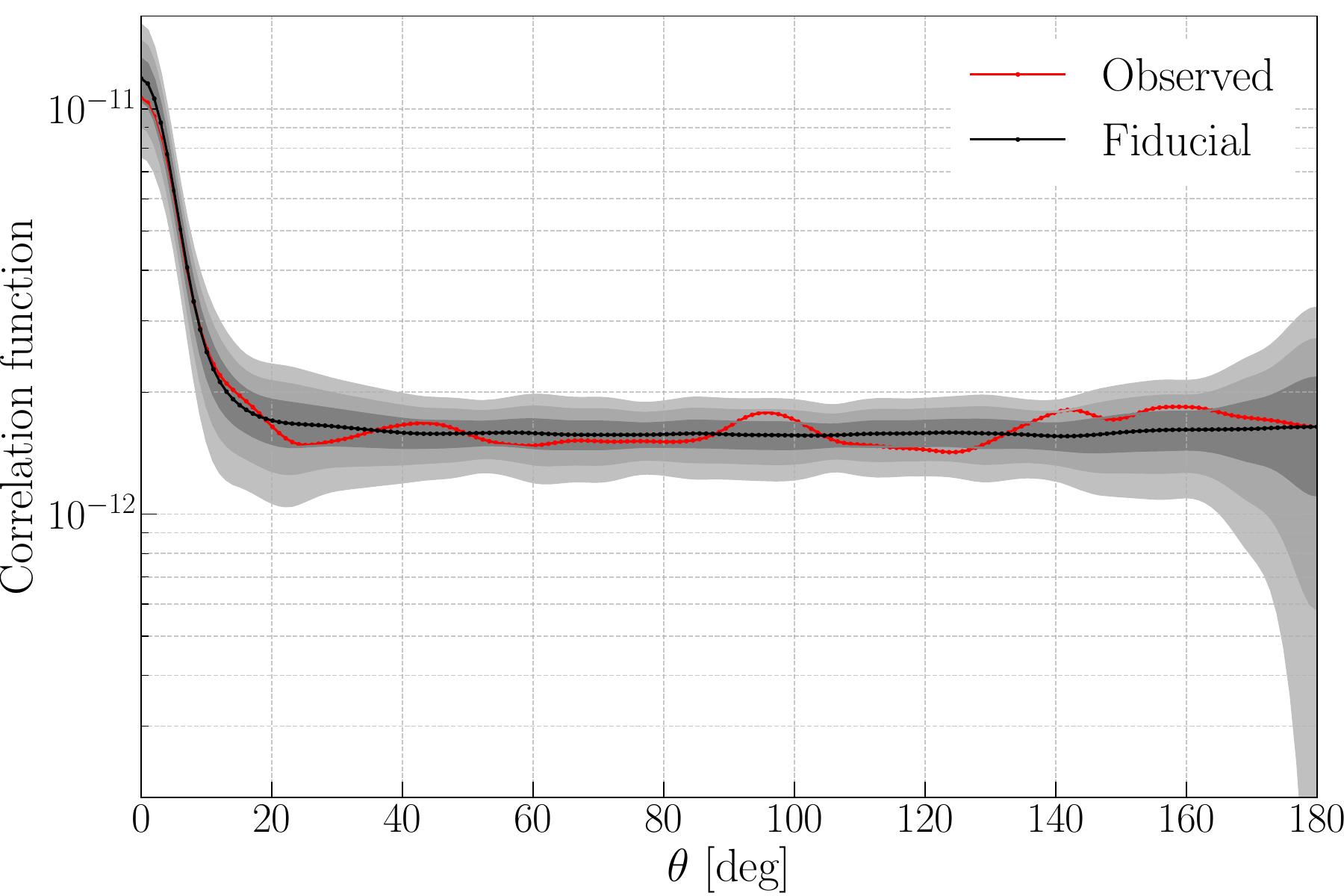}
  \captionsetup{justification=justified}
  \caption{The observed correlation function of the O3 BBH events (red curve) and the fiducial correlation function under the isotropic hypothesis (black curve). The gray-filled regions denote 1 -- 3$\sigma$ deviations from the mean.}
\label{fig:cf}
\end{figure}

Finally, we test the isotropy hypothesis with the CF. Figure \ref{fig:cf} shows the CF for the observed set and the fiducial correlation function obtained from the 100 synthetic sets under the isotropic hypothesis, where we have set the window function resolution to $\sigma_{\rm res}=l_{\rm max}$. Consistent with the power spectrum result, the observed CF is in agreement with the fiducial isotropic distribution within 2$\sigma$. The behavior of the CF at small scales, $C(\theta)=(\theta/\theta_0)^{1-\gamma}$, provides a test of isotropy \cite{Cavaglia:2020fnc}. We first compute the power-law slope $\gamma$ of each synthetic CF at the minimum angular resolution $\theta_{\rm res,s}$ with a log-log fit. Averaging the values, we obtain a fiducial value of $\gamma_{\rm s}=2.05\pm 0.35$, which is consistent with an isotropic distribution ($\gamma=2$). We then compute the power-law slope for the observed set at the same angular scale. The observed power-slope is $\gamma_{\rm o}=1.96$. This is in agreement with the null isotropic hypothesis with a p-value of 0.45.

\paragraph{\bf Conclusions---}
In this paper, we have developed a new, simulation-based inference framework to probe the spatial distribution of observed, foreground GW events. Our approach compares the power spectrum of observed GW signals to a fiducial power spectrum from a theoretical distribution.  
As an application of this method, we tested the isotropy hypothesis of the BBH mergers observed during the O3 LVK observing run. As foreseen \cite{Essick:2022slj,Namikawa:2020twf,Banagiri:2020kqd,Payne:2020pmc}, we found no evidence of anisotropy at the 2$\sigma$ confidence level. 

Our method provides a powerful framework for testing the universe's LSS that complements current GW background searches~\cite{KAGRA:2021mth, KAGRA:2021rmt}. Due to the phase-coherence of matched-filter searches employed in GWTC-3~\cite{GWTC3}, we are able to access higher multipole moments than background searches~\cite{Renzini:2021iim}.  Relying on resolved sources allows us to achieve astrometric resolution at the square degree level~\cite{Baker:2019ync}. Although the two approaches essentially target the same signal in the limit of many detections, our method has a higher resolution and is more sensitive than background analyses.

A first, straightforward extension of this work is to refine the test of BBH isotropy as more GW events are discovered. Tests of specific theoretical models of anisotropic distributions and cross-correlations with astrophysical populations in the EM domain are two additional applications. Our approach can also be directly extended to include information about the source distances. Statistical associations between the observed GW populations and other extragalactic populations may be within reach of current and next-generation GW detectors. This method will provide a means to rapidly detect and quantify any such associations.

\paragraph{\bf Acknowledgements---}
This material is based upon work supported by NSF's LIGO Laboratory which is a major facility fully funded by the National Science Foundation. The authors acknowledge computational resources provided by the LIGO Laboratory and supported by NSF Grants PHY-0757058 and PHY-0823459. M.C.~and Y.Z.~are partially supported by the U.S.\ National Science Foundation (NSF) through award PHY-2011334. N.K.~is supported by King's College London through an NMES Funded Studentship. A.I.R.~is supported by the NSF award PHY-1912594. M.S.~is supported in part by the Science and Technology Facility Council (STFC), United Kingdom, under the research grant ST/P000258/1. The authors thank Joe Romano for carefully reading the manuscript and providing useful comments. We also thank Sylvia Biscoveanu for discussions regarding PE using \texttt{bilby\_pipe}, Derek Davis for discussion on detection statistics, Stuart Anderson for helping provide computing resources, and Reed Essick for useful comments on the manuscript. This manuscript was assigned LIGO-Document number P2300106.

\bibliographystyle{unsrt}
\bibliography{references}

\end{document}